\documentclass[nofootinbib,twocolumn,prd,aps,superscriptaddress,longbibliography]{revtex4-2}

\pdfoutput=1   

\usepackage[T1]{fontenc} 
\usepackage[makeroom]{cancel}
\usepackage{physics}
\usepackage{empheq}
\usepackage{graphicx}
\usepackage{tabularx}
\usepackage{natbib}
\usepackage{amsmath}
\usepackage{dsfont}

\usepackage[
colorlinks=true,        
citecolor=blue,         
linkcolor=blue,         
urlcolor=blue           
]{hyperref}             
\usepackage{xcolor} 

\usepackage{mathrsfs}

\usepackage{enumitem}

\usepackage{lmodern}
\usepackage{txfonts}
\usepackage{color}

\usepackage{float}

\newcommand{\comment}[1]{}

\newcommand {\mathsym}[1]{{}}
\newcommand {\unicode}[1]{{}}

\newcommand{\nn}{\nonumber}

\tabularnewline

\def \be {\begin{equation}}
\def \ee {\end{equation}}
\def \bea {\begin{eqnarray}}
\def \eea {\end{eqnarray}}
\newcommand{\eq}[1]{(\ref{#1})}

\begin{document}
\title{Conic Sections on the Sky: Shadows of Linearly Superrotated Black Holes}

\author{Feng-Li Lin}
\email{fengli.lin@gmail.com}
\affiliation{Department of Physics, National Taiwan Normal University, Taipei, Taiwan 11676}

\author{Avani Patel}
\email{avani.physics@gmail.com}
\affiliation{Department of Physics, National Central University, Taoyuan, Taiwan 320317}
\affiliation{Institute of Physics, Academia Sinica, Taipei, Taiwan 11529}
\affiliation{Department of Physics, National Taiwan Normal University, Taipei, Taiwan 11676}

\author{Jason Payne}
\email{jasonpayne16@gmail.com}
\affiliation{Department of Physics, National Taiwan Normal University, Taipei, Taiwan 11676}

\date{\today}
\begin{abstract}
Soft hairs are an intrinsic infrared feature of a black hole, which may also affect near-horizon physics. In this work, we study some of the subtleties surrounding one of the primary observables with which we can study their effects in the context of Einstein's gravity: the black hole shadow. First, we clarify the singular pathology associated with black holes with soft hairs and demonstrate that the metrics of linearly superrotated black holes are free of near-zone pathologies due to appropriate asymptotic falloff conditions being imposed on the event horizon. We then analytically construct the photon orbits around such black holes, derive the shadow equation for near-zone observers, and find that the linear superrotation hairs deform the circular shadow of a bald Schwarzchild black hole into an ellipse. This is in sharp contrast to their supertranslated counterparts, which only shift the position of the center of the circular shadow but do not change its shape. Our results suggest a richness to the observable effects due to the infrared structures of Einstein's gravity. 
\end{abstract}

\maketitle

\section{Introduction}

The discovery of Hawking radiation \cite{Hawking:1975vcx} supported Bekenstein's proposal \cite{Bekenstein:1972tm, Bekenstein:1973ur} that black holes behave as thermal systems and, at the same time, asked for the microscopic origin of the area-law entropy of a generic black hole. Amongst many, one promising possibility is to attribute this to some degrees of freedom living in the near-horizon region that do not violate the no-hair theorem, i.e. do not affect asymptotic quantities such as mass, charge, and angular momentum. Within the context of Einstein's gravity, such a possibility can be realized by soft hairs \cite{Hawking:2016sgy}, which were originally conceived of as the conserved charges associated with asymptotic Bondi-Metzner-Sachs (BMS) symmetries \cite{Bondi:1962px, Sachs:1962wk, Sachs:1962zza}. 
A natural question that arises then is how one might detect such soft hairs. 

Although soft hairs do not change a black hole's mass, charge, and angular momentum, they do change its horizon structure. As such, observables relevant to the horizon structure may serve as tools for detecting soft hairs. One such observable is the black hole's shadow formed by marginally escaping light rays reaching a celestial observer. Indeed, the shadows of supermassive black holes such as M87 located at the center of our galaxy have been observed by the Event Horizon Telescope (EHT) \cite{Akiyama:2019bqs, Akiyama:2019brx, Akiyama:2019cqa, Akiyama:2019eap, Akiyama:2019fyp, Akiyama:2019sww, EventHorizonTelescope:2021dqv}. The shadow is closely related to so-called light rings \cite{Synge:1966okc, Bardeen:1973tla}, or more generally fundamental photon orbits (FPOs) \cite{Claudel:2000yi,Cunha:2017eoe, Cunha:2020azh}, which are unstable critical closed photon orbits around the black hole that serve as the dividing boundary between escaping and in-falling light rays. The shadows of Schwarzschild black holes implanted with supertranslation hairs \cite{Hawking:2016sgy, Compere:2016hzt} have been studied by both analytical \cite{Lin:2022ksb, Zhu:2022shb} and numerical ray-tracing methods \cite{Lin:2022ksb}, and these results show that the effect of supertranslation hairs is merely a shift in the center of the shadow, with no change in its shape. This implies that there is no observable imprint of supertranslation hairs on the black hole's shadow. 

Besides supertranslation hairs, a black hole can also be dressed by superrotation hairs \cite{Hawking:2016sgy, Hawking:2016msc}. These are associated with the recently discovered superrotation symmetries arising as an extension of BMS symmetries to conformal symmetries of the celestial sphere \cite{Barnich:2009se, Barnich:2010ojg, Barnich:2011mi,  Cachazo:2014fwa, Kapec:2014opa, Campiglia:2015yka}. In this work, we will study the effect of superrotation hairs on the black hole shadow and show that they can be observable. There are some subtleties, however, to this problem. Firstly, no exact solution exists for superrotated black holes; thus, we can only implant the superrotation hairs by acting on the bald Schwarzschild metric with linear gauge transformations corresponding to asymptotic isometries \cite{Strominger:2017zoo}. This method has also been applied to the linear supertranslation hairs \cite{Strominger:2017zoo, Lin:2020gva, Lin:2022ksb, Zhu:2022shb}, and the result agrees with the first-order expansion of the full supertranslated black hole metric of \cite{Compere:2016hzt}. Secondly, various pathologies are connected to the horizon structures for a BMS-type construction of linearly superrotated black holes. This will cast doubt on the validity of such metrics in the near-horizon region, which we will call the near-zone. Since the black hole shadow construction relies heavily on the photon orbits near the black hole, this doubt may hinder such a construction.  

We shall first look for (linearly) superrotated black holes with smooth near-zone metrics to bypass the above pathologies. Once such metrics are ready, the shadow construction will follow the standard procedure.  Our resolution comes from understanding the origin of the pathologies. Recall that linear soft hairs can be obtained by acting with large diffeomorphisms on the bald metric. Moreover, such gauge transformations/diffeomorphisms should be singular, as otherwise the resultant soft hairs will just be gauge artifacts and are therefore unphysical. The singular nature of such soft hair dressing on a black hole furthermore allows for new (non-spherical) solutions to bypass Birkhoff's theorem, for which the uniqueness of the spherically symmetric Schwarzschild metric should hold only for globally regular diffeomorphisms.

Thus, it is impossible to find a globally regular metric for black holes implanted with soft hairs. The situation is similar to the Dirac string singularity of a magnetic monopole in electromagnetism. Inspired by this analogy, we can try to move the singular pathology out of the near-zone via an appropriate gauge transformation. There are two null boundaries in a black hole spacetime with which one can define the asymptotic isometries for soft hairs: one is at null infinity and the other is at the event horizon. Appropriate falloff conditions will ensure regularity around either boundary. When defining the soft hairs in the Bondi gauge, the metric is regular near null infinity, which we call the far-zone, and one should expect singular pathologies in the near-zone. On the other hand, we can instead define the soft hair in the near-zone, such that the singularity will be shifted to the far-zone, resulting in a non-asymptotically flat metric. 
Indeed, such linearly supertranslated and superrotated black holes have been studied in \cite{ Donnay:2015abr, OzEling:2016JHEP}. With a metric that is smooth in the near-zone, we can construct the light rings, or FPOs, free of any pathologies, which are then valid for constructing black hole shadows with respect to some near-zone observers.  This will be the black hole metric with linearly superrotated hairs that we adopt to construct the shadow for near-zone observers. We will demonstrate that there is a non-trivial deformation effect of linear superrotation hairs on the shape of the shadow.  

 The remainder of this paper is organized as follows. In the next section, we will consider linear superrotations in both the far- and near-zones. We then write down the near-zone linearly superrotated black hole metric relevant for constructing the corresponding shadow and discuss their physical implantation via shockwaves. In section \ref{sec3}, we will analytically construct the shadows for the linearly superrotated black holes by first considering the FPOs and then deriving the analytic form of the shadow equation, which turns out to that of an ellipse. Finally, we conclude the paper in section \ref{sec4}.


\section{A Tale of Two Superrotations}

Unlike the hard hairs of black holes, which are associated with isometries of the underlying metric, soft hairs are associated with asymptotic isometries defined by appropriate falloff conditions of the metric components. However, there are two asymptotic boundaries in a black hole background: one is in the far-zone region near null infinity, and the other is in the near-zone region near the black hole's event horizon. Given a known exact solution for a black hole with soft hairs, such as the supertranslated black holes constructed in \cite{Compere:2016hzt}, one may be able to define these soft hairs globally. Otherwise, we can only construct the soft hairs either at null infinity or at the event horizon, and relate them via an appropriate coordinate transformation. This is the case for the superrotation soft hairs considered in this work. 

When considering the detection of soft hairs by observing their effects on the black hole shadow, one may ask in which region we shall consider the soft hairs for this purpose. As the black hole shadow inherently reflects photon orbits in the near-zone region, it seems better to construct soft hairs in this region. On the other hand, the soft hairs associated with the Bondi-Metzner-Sachs (BMS) symmetries near null infinity are well studied \cite{Bondi:1962px, Sachs:1962wk, Sachs:1962zza, Strominger:2017zoo}, and so it is also necessary to understand their implications for the construction of black hole shadows. In this section, we will discuss superrotation hairs in both near-zone and far-zone regions for this purpose. The metric of such black holes can be generated by performing a large gauge transformation on the following bald Schwarzschild metric
\be
    \bar{g}_{\rho\sigma} dx^{\rho} dx^{\sigma}=-\left(1-\frac{2M}{r}\right) dv^2 + 2 dv dr + r^2 {\gamma}_{AB} dx^A dx^B \,,\label{bald_metric}
\ee
where $x^A=(z,\bar{z})$ are the stereographic coordinates of the celestial $2$-sphere with $\gamma_{z\bar{z}}=\gamma_{\bar{z}z}={2/(1+z\bar{z})^2}$ defining the nonzero components of $\gamma_{AB}$. 
Since this metric is an exact solution of Einstein's equation, it can work for both the near- and far-zone regions.

\subsection{Far-zone superrotations}
We first consider the far-zone region, in  which linear superrotation hairs are generated by the following superrotation vector \cite{Strominger:2017zoo, Barnich:2009se, Barnich:2011mi}, 
\begin{equation}
\zeta_Y=\frac{D\cdot Y}{2}\big[(v-r) \partial_r+v \partial_v \big] +\big[Y^A+\frac{v}{2r}D^{A}D\cdot Y \big] \partial_{A} ,\label{SR_v}  
\end{equation}
where $D_A$ is the covariant derivative with respect to $\gamma_{AB}$, and $Y^A(x^B)$ are the conformal Killing vectors (CKVs) of the sphere satisfying 
\be
D_A Y_B + D_B Y_A = \gamma_{AB} D\cdot Y
\ee
with $D\cdot Y:= D_A Y^A$. The CKV conditions imply that the $Y^A$ are holomorphic (or even meromorphic), i.e., $Y^z=Y^z(z)$ and $Y^{\bar{z}}=Y^{\bar{z}}(\bar{z})$. One can then generate the linearly superrotated black hole metric $g_{\rho\sigma}$ by
\be
g_{\rho\sigma} =\Big( 1+ {\cal L}_{\zeta_Y} \Big) \bar{g}_{\rho\sigma}\;,
\ee
resulting in
\cite{Strominger:2017zoo}:
\begin{eqnarray}
    g_{\rho\sigma} dx^{\rho} dx^{\sigma}&=&\left[-\left(1-\frac{2M}{r}\right)+\frac{3 M}{r} D\cdot Y \right] dv^2  \nonumber \\
&& +  2 dv dr +  r^2 \left(\gamma_{AB}+ \kappa_{AB}\right)d\Theta^Ad\Theta^B \label{metricH}
\end{eqnarray}
where  $\Theta^A:=(\theta,\phi)$ are the typical spherical coordinates on the 2-sphere, and 
\be 
    \kappa_{AB}= -\left(1-\frac{v}{r}\right)D\cdot Y \gamma_{AB}+ 2 D_{(A|}\left[Y^C+\frac{v}{2r} \left(D^CD\cdot Y\right)\right] \gamma_{C|B)}\;.
\ee
It is easy to see that the metric preserves the Bondi gauge when the $Y^A$ are CKVs \cite{Strominger:2017zoo}, but the metric is no longer static. As the Bondi gauge is relevant for far-zone asymptotic isometries, we may expect the metric to be pathological in the near-zone region. This is because the diffeomorphism \eq{SR_v} should be singular in order for the superrotation to be physical. Otherwise, it will just produce a gauge artifact. Indeed, besides the event horizon at $r= 2M + 3 M D\cdot Y$,  there is a superhorizon for which ${\rm det}[ g_{\rho\sigma}]=0$ at $r=r_{\rm SH}$ with
\be
r_{\rm SH}=- v \Big[\partial_{\theta} D^{\theta} + \cot\theta D^{\theta} + \partial_{\phi} D^{\phi} + 2  \Big] \big(D\cdot Y \big) \;. \label{rSH}
\ee
Note that $r_{\rm SH}$ grows pathologically with $v$. Moreover, weak cosmic censorship will be violated if 
\be
3 |D\cdot Y| > 2. 
\ee
These pathologies of the metric \eq{metricH}, including its linear growth in $v$, cast doubt on adopting this metric for consideration of the photon orbits in the near-zone region, and therefore for constructing black hole shadows as well.

\subsection{Near-zone superrotations}

One can immediately see that all of the pathologies of \eq{metricH} discussed earlier can be lifted simultaneously if we generalize the $Y^A$ to not be CKVs, but rather satisfying $D\cdot Y=0$ \footnote{It is easy to see that holomorphic (or meromorphic) CKVs $Y^A$ are not compatible with $D\cdot Y=0$ unless $Y^A=0.$}. In such cases, the metric is reduced to
\be \label{sr_met_DY_1}
 g_{\rho\sigma} dx^{\rho} dx^{\sigma}= \bar{g}_{\rho\sigma} dx^{\rho} dx^{\sigma} + r^2 (D_A Y_B+ D_B Y_A) dx^A dx^B\;, 
\ee 
which differs from the bald metric by just a shape-deformed celestial $2$-sphere. Since these $Y^A$ are not CKVs, the metric \eq{sr_met_DY_1} is no longer in the Bondi gauge, as expected. This implies that we cannot interpret it as a black hole with BMS-type soft hairs. Despite this, the shape-deformed $2$-sphere suggests that the metric should carry some soft-hair information. Thus, we may wonder if the metric can be interpreted as relating to superrotation hairs outside of the far-zone. Indeed, they are superrotation hairs as seen from the near-zone region \cite{Donnay:2015abr, OzEling:2016JHEP}.

To realize this, we follow the construction of \cite{OzEling:2016JHEP}. To emphasize the near-zone nature of the construction, we define a new coordinate $\rho=r-2M$ so that \eq{bald_metric} takes the form of the so-called Gaussian null coordinates, and we expand the metric to the order of $\rho$. To define the falloff conditions for the asymptotic isometries for the near-zone soft hairs, we shall perform the gauge transformation ${\cal L}_{\xi}$, which will be constrained by requiring that the metric maintains the Gaussian null coordinates, i.e.,  \begin{equation}\label{}
    \mathcal{L}_\xi g_{v\rho}=\mathcal{L}_\xi g_{\rho\rho}=0,
\end{equation}
as well as preserving the horizon \footnote{There are some differences in \cite{Donnay:2015abr} and \cite{OzEling:2016JHEP} regarding the falloff conditions at sub-leading order, but these will not affect our leading-order results. In \cite{Donnay:2015abr} they moreover go on to explicitly identify the full BMS-like algebra corresponding to these near-horizon supertranslations and superrotations.} in the sense 
\begin{equation}
    \mathcal{L}_\xi g_{vv}=\mathcal{L}_\xi g_{vA}=0+\mathcal{O}(\rho)\;.
\end{equation}

As was seen in \cite{OzEling:2016JHEP}, the required $\xi$ takes the explicit form
\begin{align} \label{xi_A}
    \xi &= T \partial_v+\big(R^A -\rho\gamma^{AB}\partial_B T \big)\partial_A + \rho\partial_v T \partial_\rho\;,
\end{align}
where the functions $T(v,x^A)$ and $R^A(x^B)$ characterize the supertranslations and superrotations in the near-zone region, respectively. We can further decompose $R^A$ by
\be
R^A=\varepsilon^{AB}\partial_B f+\partial^A g,\label{Rdecomp}
\ee
for arbitrary functions $f$ and $g$ on the $2$-sphere. By comparing \eq{xi_A} with \eq{SR_v}, we can conclude that 
\be\label{xi_zeta}
\xi\vert_{T=0, f=0} = \zeta_Y\vert_{D\cdot Y=0} \;.
\ee
with 
\be
Y^A=D^A g
\ee
in the near-zone region. Then, $D\cdot Y=0$ further implies
\be
D^2 g=0
\ee
which means we are dealing with an area-preserving diffeomorphism as in \cite{OzEling:2016JHEP}.

The equivalence relation \eq{xi_zeta} implies that the diffeomorphism $\zeta_Y\vert_{D\cdot Y=0}$ can be treated as a large gauge transformation generating the near-zone superrotation hairs. On the other hand, if the $Y^A$ are CKVs, it can be treated as that of the far-zone superrotation hairs. The convenience in this lies in the fact that the bald metric \eq{bald_metric} is an exact solution, which can then be used to study both the near- and far-zone asymptotic isometries depending on the choice of the diffeomorphisms with their corresponding falloff conditions.

Moreover, for a bald metric with nonzero $g_{vA}$ taking the form
\be
g_{vA}=\varepsilon_A^B \partial_B \psi + \partial_A \varphi\;,
\ee
the superrotation charge generated by \eq{Rdecomp} is given by \cite{OzEling:2016JHEP}
\begin{equation}
    Q_{\rm sr}=\int d^2x\sqrt{\gamma}\big(fD^2\psi+gD^2\varphi\big)\;. \label{srcharge}
\end{equation}
However, it is zero for the Schwarzschild metric with $g_{vA}=0$.

\subsection{Black hole metric with near-zone linear superrotation hairs for the shadow construction}

In the previous discussion, we have considered superrotation hairs in stereographic coordinates on the $2$-sphere. It is, however, more convenient to adopt the usual spherical coordinates $\Theta^A$ for the construction of black hole shadows. Accordingly, we adopt the convention in this work that $Y^{\theta}(\theta,\phi)$ parameterize the near-zone superrotation hairs and impose the $D\cdot Y =0$ condition. This can be solved for $Y^{\phi}$ by
\be\label{DY_f}
Y^{\phi} = -\int d\phi \big[ \partial_{\theta} Y^{\theta} + \cot\theta Y^{\theta} \big] \quad {\rm with} \quad || Y^{\theta} || =1
\ee
for a given $Y^{\theta}$ with the maximum of its absolute value denoted by $||Y^{\theta}||$. Note that we have set $|| Y^{\theta}||=1$ and introduced a small parameter $\epsilon$ capturing the overall size of the superrotation hair in what follows. In this way, the metric \eq{sr_met_DY_1} transforms into the following form\footnote{Note that this metric for a linearly superrotated black hole with $D\cdot Y=0$ is different from the one of a slowly rotating black hole, which takes the following form:
\be 
ds^2_{\rm slow rot}= \bar{g}_{\rho\sigma} dx^{\rho} dx^{\sigma} - 2 a r^2 \sin^2\theta dv d\phi\;. \nonumber 
\ee
The hairy term in \eq{sr_met_DY} mainly deforms the 2-sphere. On the other hand,  the last term in the above metric introduces nonzero $g_{v\phi}$.},
\bea \label{sr_met_DY}
 g_{\rho\sigma} dx^{\rho} dx^{\sigma} &=& -\left(1-\frac{2M}{r}\right) dv^2 + 2 dv dr + r^2 (d\theta^2 + \sin^2\theta d\phi^2) 
 \nn \\
 && + \epsilon r^2 (D_A Y_B+ D_B Y_A) d\Theta^A d\Theta^B\;.
\eea 
Along with the condition \eq{DY_f}, this is the metric we will adopt for constructing the shadows of a black hole implanted with near-zone linear superrotation hairs.   

Although the metric \eq{sr_met_DY} is free of horizon pathologies, it is not asymptotically flat as it holds only in the near-zone region. This is the price we pay by shifting the near-zone singularity to the far-zone when adopting the near-zone falloff conditions for asymptotic isometries. However, we shall consider this unavoidable singularity as physical evidence for the soft hairs. That is, even formally, the soft hairs can be generated by (large) gauge transformations; they are not gauge artifacts because these gauge transformations are singular. This is much like the magnetic monopole of electromagnetism, for which the associated Dirac string singularity can be moved around by gauge transformations. Similar arguments for supertranslation hairs can also be found in \cite{Hawking:2016msc, Pasterski:2020xvn}.

Thus, when constructing the shadows seen by a celestial observer, we should remember that the separation between the black hole and the observer should be within the valid range for the metric \eq{sr_met_DY}. Despite such limitations on this metric, it is already good enough to study the effects of linear superrotation hairs on the shape of the shadow for some observers with a reasonable separation from the black hole. 

\subsection{Implanting linear superrotation hairs}
Another natural concern one might have is how to physically implant such superrotation hairs. This can be addressed in much the same way as in the supertranslation case \cite{Hawking:2016sgy}. One can imagine that our metric \eq{sr_met_DY} arises from a shockwave arriving from outside the near-zone and colliding with our bald black hole at advanced time $v_0$. We find that this is indeed the case for a shockwave with a stress-energy profile of the form
\begin{subequations}
\begin{align}
    T_{v\theta} &= \frac{\epsilon\delta(v-v_0)}{2}\bigg(-2\csc\theta\sec\theta\partial_\phi Y^\phi+\csc^2\theta\partial_\phi^2 Y^\theta \nonumber \\
    &\hspace{.7in}+2(\cot\theta-\csc\theta\sec\theta)\partial_\theta Y^\theta-\partial_\theta\partial_\phi Y^\phi\bigg) \\
    T_{v\phi} &= \frac{\epsilon\delta(v-v_0)}{2}\bigg(\cot\theta\partial_\phi Y^\theta+3\cos\theta\sin\theta\partial_\theta Y^\phi \nonumber \\
    &\hspace{1.4in}-\partial_\theta\partial_\phi Y^\theta+\sin^2\theta\partial_\theta^2 Y^\phi\bigg) \\
    T_{\theta\theta} &= -2\epsilon r\delta(v-v_0)\partial_\theta Y^\theta \\
    T_{\theta\phi} &= -\epsilon r\delta(v-v_0)\big(\partial_\phi Y^\theta+\sin^2\theta\partial_\theta Y^\phi\big) \\
    T_{\phi\phi} &= 2\epsilon r\delta(v-v_0)\sin^2\theta\partial_\theta Y^\theta
\end{align}
\end{subequations}
By solving the linearized Einstein equations with such a profile, one indeed recovers a metric of the form
\begin{equation}
    g_{\rho\sigma} = \bar{g}_{\rho\sigma} +\theta(v-v_0)\delta g_{\rho\sigma}
\end{equation}
with $\delta g_{\rho \sigma} dx^{\rho} dx^{\sigma}$ given by the second line of \eq{sr_met_DY}, describing a bald black hole implanted with the near-zone linear superrotation hairs at $v_0$. Note that the energy flux of this shock wave depends on its radial position, but is finite within the near-zone region. The scale of the stress tensor of the shock wave, which determines the scale of the soft hair, should highly depend on the surrounding environment at the early stage of the black hole's formation. Thus, the overall scale of the superrotation hairs will depend on the activity strength of the nearby matter surrounding the black hole.

\section{The Analytic Form of the Shadow}\label{sec3}

The shadow of a black hole arises from the critical photon orbits, below which the light will fall into the black hole and not be observed. For Schwarzschild black holes, these critical photon orbits are planar, circular, and called light rings (LRs) \cite{Bardeen:1973tla}. For more generic black holes, such as Kerr black holes, they can be non-planar and non-circular and are called fundamental photon orbits (FPOs) \cite{Claudel:2000yi, Cunha:2017eoe, Cunha:2020azh}. For more general and complicated metrics, such as \eq{metricH}, the FPOs may not even exist. In such cases, one can only adopt the numerical ray-tracing method to determine the photon orbits for the shadow construction. However, in this work, we will consider only near-zone linear superrotation hairs with the metric given by \eq{sr_met_DY}, which differs from the bald metric by just area-preserving deformed terms on the $2$-sphere. This metric no longer has any near-zone pathology, with the event horizon still at $r=2M$. Thus, we can determine the associated light rings or FPOs straightforwardly, which in turn allows for the analytical derivation of the shape of the shadow.

\subsection{Fundamental Photon Orbits}
We start with a photon orbit in the bald metric, which is expressed by the Hamilton-Jacobi method in terms of the canonical momentum, denoted by $\bar{p}^{\mu}$. The latter is determined by $x^{\mu}$ and the constants of motion, i.e.,
\be
\bar{p}^{\mu}=\bar{p}^{\mu}(x^{\mu}; E, \lambda, \kappa)
\ee
where $E$, $\lambda$, and $\kappa$ are the constants of motion corresponding to the energy, reduced azimuthal angular momentum, and the Carter constant, respectively.  Once the superrotation hair turns on, the geodesic equations in the new frame defined by \eq{sr_met_DY} can be obtained by an active coordinate transformation, i.e., 
\be\label{pH1}
p^{\mu}(x^{\mu}; E, \lambda, \kappa)= \bar{p}^{\mu}(x^{\mu}; E, \lambda, \kappa) + \epsilon {\cal L}_{\zeta_Y} \bar{p}^{\mu}(x^{\mu}; E, \lambda, \kappa)\;.
\ee
Explicitly, they are 
\begin{subequations}\label{pH_f}
\begin{align}
p^v &= E \Big(\frac{r}{r -2 M}\Big) \Big(1+ {\cal R}\Big)\;, \label{pB1}
\\
p^r &= E {\cal R}\;,  \label{pB2}
\\ 
p^{\theta} &= \frac{E}{r^2} {\cal K}+\frac{\epsilon E}{r^2} \Big[ \csc^2\theta\; \bigg(\lambda^2 \frac{\cot\theta}{{\cal K}} Y^{\theta} -\lambda \partial_{\phi}Y^{\theta} \bigg) -{\cal K} \partial_{\theta} Y^{\theta}  \Big] \;,\label{pB3}
\\
p^{\phi} &= \frac{E\lambda}{r^2\sin^2\theta} 
+ \frac{\epsilon E}{r^2} \Big[\lambda \csc^2\theta \big(2 \cot\theta \; Y^{\theta} + \lambda \partial_{\phi}Y^{\phi} \big) + {\cal K} \partial_{\theta} Y^{\phi}  \Big] 
\;, 
\label{pB4}
\end{align}
\end{subequations}
where in each equation, the first term corresponds to that of the bald metric and 
\be
{\cal R}:=\sqrt{\frac{r^3-r \kappa + 2 M \kappa}{r^3}}, \qquad {\cal K}:=\sqrt{\kappa - \lambda^2  \csc^2\theta} \;.\label{R-K}
\ee
It is important to remember that the above constants of motion are \textit{merely labels inherited from the photon momenta of the bald metric}, but are not themselves constants of motion of the superrotated spacetime. We can then view the photon momenta given by \eq{pH_f} as deformations of the bald momenta, which are still labeled by these inherited constants of motion. Alternatively, one could take the passive approach and consider the geodesic equation in this new frame; however, we will not have the inherited constants of motion to label the photon orbits. As these are central to our later analytic construction of the shadow, the passive approach will not be considered here. 


We furthermore note that since the $v$- and $r$-components of the photon momenta, \eq{pB1} and \eq{pB2}, are the same as the ones in bald metric, the FPO defined by $\dot{r}=\ddot{r}=0$  (or $p^r=\dot{p^r}=0$ where the over-dot denotes the derivative with respect to the worldline time $t$) has the same radius $r=3 M$ as the light ring in the bald metric, with $\kappa=27 M^2$. However, these FPOs have more complicated dynamics along the $\theta$- and $\phi$-directions, i.e., they no longer lie in the equatorial plane. They are deformed light rings obtained from the corresponding orbits for the bald metric, labeled by $E$ and $\lambda$, but with $\kappa= 27 M^2$. From now on, we will only consider
\be
\kappa= 27 M^2 \quad {\rm for \; light \;  rings \; or \; FPOs.} 
\ee

One can obtain the FPOs by directly solving the geodesic equations of \eq{pH_f} perturbatively in $\epsilon$, up to ${\cal O}(\epsilon)$. Then, these FPOs will be labeled by the inherited constants of motion appearing in \eq{pH_f}. On the other hand, we can also exploit the covariance of the geodesic equations and their solutions by shifting the light ring in the bald metric, i.e., $\bar{x}^{\mu}_{\rm LR}=(3 E t, 3 M, \frac{\pi}{2}, -\frac{E \lambda}{9 M^2}t)$ parameterized by the worldline time $t$, with the superrotation vector  $- \zeta_Y|_{D\cdot Y=0}$. The result is 
\be\label{deformed_LR}
\bar{x}^{\mu}_{\rm LR} \rightarrow \bar{x}^{\mu}_{\rm LR} - \Big(0,0, \epsilon Y^{\theta}(\bar{x}^{\mu}_{\rm LR}), \epsilon Y^{\phi}(\bar{x}^{\mu}_{\rm LR}) \Big)\;,
\ee
which turns out to be the same as the FPOs obtained from the geodesic equations \eq{pH_f}. This is not surprising because the metric \eq{sr_met_DY} is regular in the near-zone region so that any diffeomorphism will preserve the critical nature of FPOs in this region. Besides, we see that such FPOs have a radius of $3M$ but with nontrivial dynamics along $\theta$- and $\phi$-directions through $\phi(t)=-\frac{E \lambda}{9 M^2}t$ contained in $Y^A(\bar{x}^{\mu}_{\rm LR})$ \footnote{Similar correspondence can be found in the case of linear supertranslation hairs characterized by the coordinate transformation:
\be
x^{\mu}\rightarrow x^{\mu} - \epsilon \xi_T \qquad \mbox{with} \qquad \xi_T:=T\partial_v -{\tfrac{1}{2}} D^2T \partial_r + {\tfrac{1}{r}} D^AT \partial_A\;, 
\ee
for an arbitrary function $T=T(\theta,\phi)$. If $T=T(\theta)$, then $E$ and $\lambda$ are still constants of motion for a photon orbit in the hairy background metric with supertranslation hairs, and the radius of the light ring can be obtained by solving the geodesic equations coupled with the null condition $g_{\mu\nu}\dot{x}^{\mu} \dot{x}^{\nu}=0$. The result is \cite{Sarkar:2022prd}
\be
r_{\rm LR}=3 M+ \frac{\epsilon}{2} T''\bigg(\frac{\pi}{2}\bigg) \;.  
\ee
This matches with $3M-\epsilon \xi^r_T\big\vert_{x=\bar{x}^{\mu}_{\rm LR}}$ from the linear supertranslation of $\bar{x}^{\mu}_{\rm LR}=(3 E t, 3 M, \frac{\pi}{2}, -\frac{E \lambda}{9 M^2} t)$ of the bald metric light ring.}.
For generic $Y^A$, the light ring wobbles about the equatorial light ring of the bald metric \footnote{Note that the integration constant $h(\theta)$ obtained by solving $D\cdot Y=0$ for $Y^{\phi}$ will not affect the FPO, as it just shifts $\bar{\phi}_{\rm LR}(t)$ by the constant $h(\tfrac{\pi}{2})$.}, as shown in Fig. \ref{fig1} below.

\begin{figure}[H]
    \centering
\includegraphics[scale=.278]{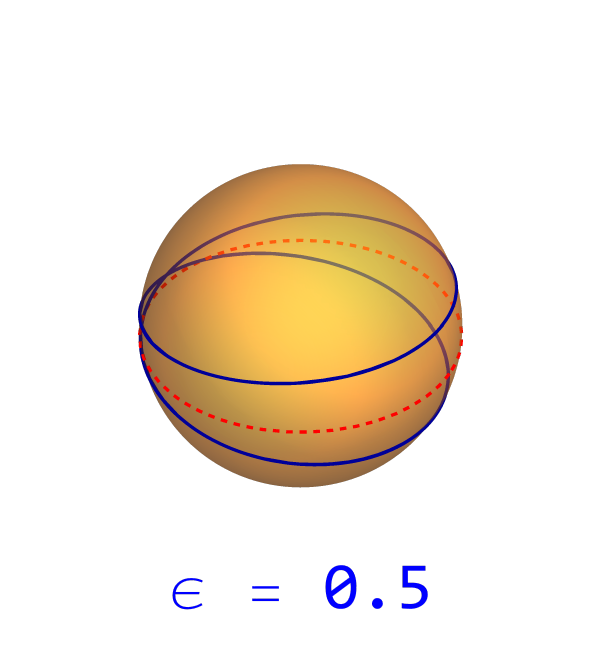}
\includegraphics[scale=.278]{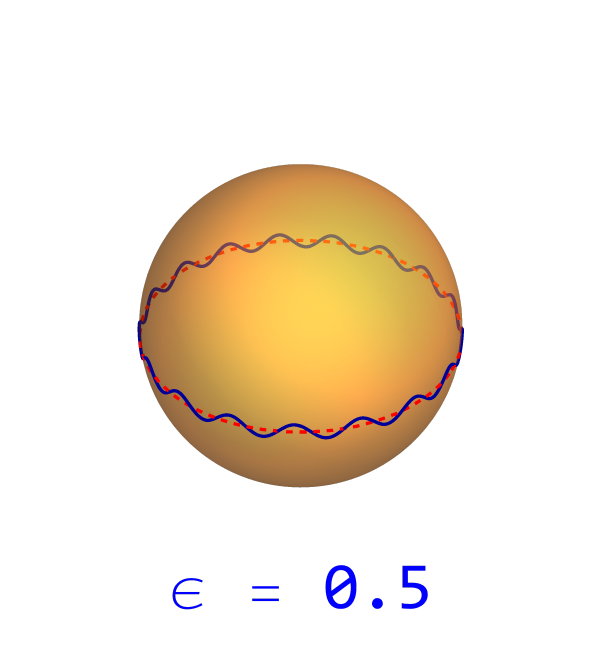}
\includegraphics[scale=.278]{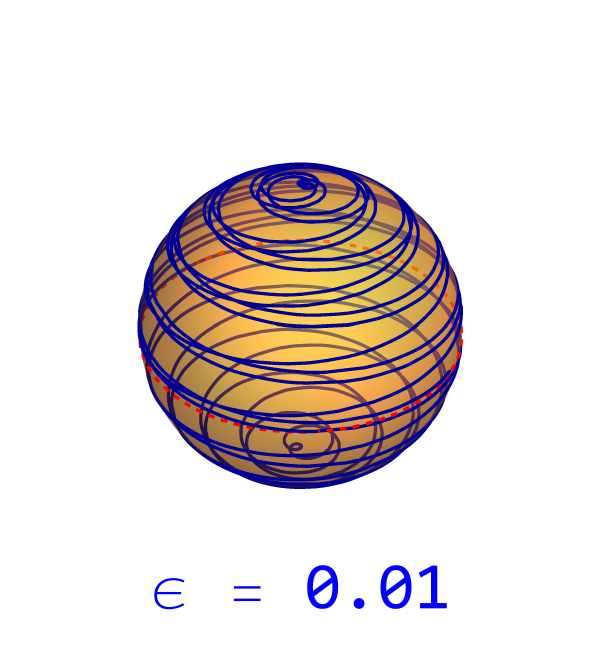}
    \caption{Illustrative examples of some of the broad features of deformed light ring/fundamental photon orbits (blue curve) for the near-zone metric \eq{sr_met_DY} compared to its circular bald counterpart (red), each labeled by the same set of constants of motion, especially $\kappa=27M^2$. For sinusoidal-type choices of $Y^\theta$ (left, $Y^\theta=\sin\tfrac{\theta}{2}\cos\frac{\phi}{2})$, the deformed light ring wraps around above and then below the bald light ring a number of times. For real spherical harmonics (center, $Y^\theta=Y^{17}_{25}$), the deformed light ring oscillates around the bald light ring, generally more prominently for larger $|\ell|$ and $|m|$. Lastly, the addition of a polynomial in $\theta$ and $\phi$ (right, $Y^\theta=Y^{17}_{25}+\theta-5\phi$) can stretch the deformed orbit more drastically around the sphere of radius $3M$. The value of $\epsilon$ is given for each case and is chosen here simply for convenience, beyond being small relative to the normalization of $Y^\theta$. In reality, it shall depend on the energy scale of the matter in the surrounding environment when implanting the superrotation hair. In each of these $Y^\phi$ determined by \eq{DY_f} and $E=L_z=M=1$.}
    \label{fig1}
\end{figure}

\subsection{Conical Shadows}
We now derive the analytic form of the shadow of black holes implanted with these superrotation hairs. Constructing shadows for Kerr black holes was first considered in  \cite{Synge:1966okc, Bardeen:1973tla, Cunningham1972TheOA} by tracing the light from the observer to the source backward. Later, a variety of constructions with the different choices of the azimuthal axis of the celestial observer's coordinate system arose, e.g., \cite{Vazquez:2003zm, Grenzebach:2014fha, Cunha:2016bpi}. See also the review \cite{Perlick:2021aok}. Thus, the resultant shadows obtained via these various methods are related by some rotation of the coordinate axes. Our preference, which we adopt here, is the method of \cite{Vazquez:2003zm}, which is briefly sketched below. 

\begin{figure}[H]
    \centering
    \includegraphics[scale=.42]{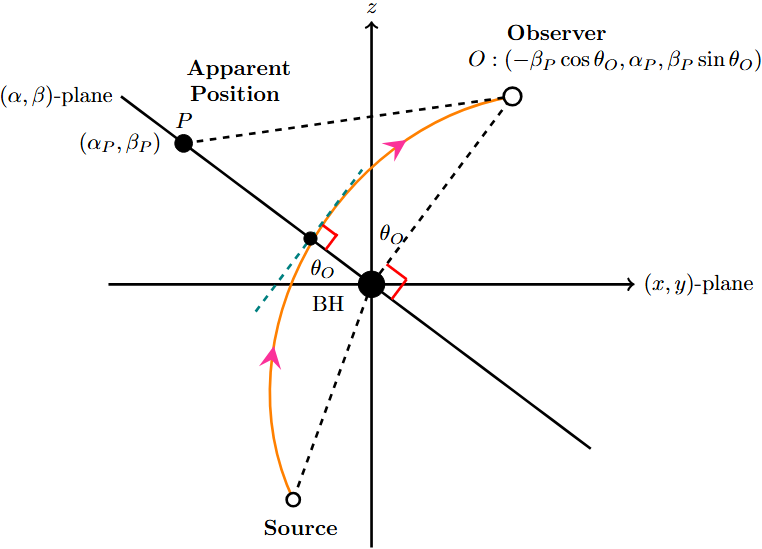}
    \caption{A marginal light ray (orange) traveling from the source around the black hole to the observer at $O$. The (approximately) flat coordinate system $(x,y,z)$, valid within the near-zone, is given. The celestial plane coordinated by $(\alpha,\beta)$ is perpendicular to the line connecting $O$ to the black hole and we have chosen for the sake of convenience to align $y=\alpha$. The intersection of the tangent to the light ray at $O$ and the celestial plane determines the apparent position $P=(\alpha_P,\beta_P)$ of the ray from the observer's perspective.}
    \label{fig:celestial}
\end{figure}

The light rays that can marginally escape from the black hole form the boundary of the shadow on the observer's celestial plane. By tracing these marginal light rays backward from the observer to the source, the boundary of the shadow is then obtained by collecting their intersections with the celestial plane. The celestial plane is perpendicular to the line connecting the observer to the black hole and almost to the marginal light rays. In Fig. \ref{fig:celestial}, we set up the coordinates of the observer $O$ and the celestial plane, which are denoted by $\vec{r}=(x,y,z)=(r\sin\theta\cos\phi, r\sin\theta\sin\phi,r\cos\theta)$ and $(\alpha,\beta)$, respectively. For simplicity, we assume the $\alpha$-axis coincides with the $y$-axis. Without loss of generality, we choose the location of $O$ at $\vec{r}_O:=(r_O \sin\theta_O, 0, r_O \cos\theta_O)$, i.e., $\phi_O=0$, and $r_O$ is the radial distance from the black hole located at the origin. 

Consider a specific light ray crossing the celestial plane with the tangent vector of its parametric curve at $O$ given by ${d\vec{r} \over dr}\big\vert_O={dx \over dr}\big\vert_O \hat{x}+ {dy \over dr}\big\vert_O \hat{y}+ {dz \over dr}\big\vert_O \hat{z}$. Denote the apparent position $P$ of this marginal light ray on the celestial plane by $(\alpha_P, \beta_P)$, which in $O$'s coordinates is given by $\vec{r}_P:=(-\beta_P \cos\theta_O, \alpha_P, \beta_P \sin\theta_O)$, as read from Fig. \ref{fig:celestial}. If the curvature effect is negligible, the curve $\overline{OP}$ in Fig. \ref{fig:celestial} is a straight line described by
\be\label{ray-tracing}
\vec{r}_P \simeq - {d\vec{r} \over dr}\Big\vert_O r_O + \vec{r}_O\;, 
\ee
where the minus sign indicates that we are performing \textit{backward} ray tracing, and the approximate equality indicates that any curvature effect on line $\overline{OP}$ is neglected. The $y$- and $z$-components of \eq{ray-tracing}, respectively, are given by
\bea\label{sh_1}
\alpha_P \simeq - r^2_O \sin\theta_O {d \phi \over dr}\Big\vert_O\;, 
\qquad \beta_P \simeq r^2_O   {d \theta \over dr}\Big\vert_O\;.  
\eea
Lastly, using the fact that $\vec{p}={d \vec{r} \over dt}$ for a light ray trajectory $\vec{r}(t)$, one finds that \eq{sh_1} yields the following equations for the shape of the shadow on the celestial plane \cite{Zhu:2022shb}
\be
\alpha \simeq -r_O^2 \sin{\theta}_O\left.\frac{p^{\phi}}{p^r}\right|_O \;, \qquad \beta \simeq r_O^2 \left.\frac{p^{\theta}}{p^r}\right|_O.
\label{alpha-beta}
\ee
Here $p^{\mu}$ is given by \eq{pH_f}. By taking the large $r_O$ limit when evaluating $p^{\mu}$ in \eq{pH_f} for the Schwarzschild black hole, i.e., $\epsilon=0$, \eq{alpha-beta} one arrives at
\be\label{circle}
\lambda = -\alpha \sin\theta_O\;, \qquad \kappa = \alpha^2 + \beta^2\;. 
\ee
The second equation indicates that the shape of the shadow for a Schwarzschild black hole is a circle with a radius of $\sqrt{\kappa}$, as is well known from a variety of constructions\footnote{The resultant shadow equation \eq{circle} is slightly different from the one given in \cite{Bardeen:1973tla, Cunningham1972TheOA} due to a different convention for the Carter constant and the choice of azimuthal axis. However, our resultant circle equation is the same as the one in \cite{Perlick:2021aok}. Despite the difference, they all describe the circular shape of Schwarzschild shadow, which is physical and coordinate-independent.} \cite{Perlick:2021aok}. Moreover, in \cite{Zhu:2022shb} \eq{alpha-beta} is used to demonstrate the circular shape of the shadow for a linearly supertranslated black hole, just as was obtained via the numerical ray-tracing method in \cite{Lin:2022ksb}. We will apply \eq{alpha-beta} to obtain the shadow for the linearly superrotated black hole considered in this paper. 

Before carrying out the explicit construction, we should first discuss the applicability of \eq{alpha-beta} to the near-zone metric 
\eq{sr_met_DY} of the linearly superrotated black hole. The approximate equality in \eq{alpha-beta} reminds us that it holds exactly only in the $r_O \rightarrow \infty$ limit. Similarly, to obtain the circular shape of the shadow in \eq{circle}, one also needs to take the same limit. However, the near-zone metric \eq{sr_met_DY} is not asymptotically flat; thus, one cannot naively apply \eq{alpha-beta} to construct the shadow for the near-zone metric \eq{sr_met_DY}. In the near-zone approach, the singularity due to the soft hairs has been shifted to the far-zone. Thus, we can assume the near-zone metric should be valid up to the middle zone of scale $r_{\rm mid}$, which should be small compared to the far-zone scale near $r=\infty$. Thus, we can define the near-zone observer by the following hierarchical relation
\be
r_S:=2M \ll r_O \le r_{\rm mid} \ll r=\infty.
\ee
Constraining $r_O$ by this relation, we can correct the shape of the shadow obtained by \eq{alpha-beta} with the subleading corrections expressed by
\be\label{formal_O}
\kappa = \alpha^2 + \beta^2 + \epsilon \delta_0(\alpha,\beta; T, Y^A) +  \sum_{n=1} \Big({r_S \over r_O}\Big)^n \delta_n(\alpha, \beta)\;.
\ee
Here, $\delta_0$ is the correction due to the linear soft hairs specified by $T$ and $Y^A$ by naive application of \eq{alpha-beta}, which can be understood to be approximately well-behaved if $r_S\ll r_O \le r_{\rm mid}$. On the other hand, the $\delta_{n\ge 1}$'s are independent of the soft hairs and can be determined order-by-order systematically for a given bald metric. For the supertranslation case, it was shown $\delta_0=0$ \cite{Lin:2022ksb, Zhu:2022shb}, but in the following, we will show that $\delta_0 \ne 0$ for the superrotation case.

Based on \eq{formal_O}, we can choose appropriate values for $\epsilon$ and $r_O$ to either suppress the contributions of $\delta_{n\ge 1}$'s, or incorporate these corrections order-by-order systematically. In either case, the effect of nonzero $\delta_0$ will come into play in the resultant shadow. In this work, we would like to emphasize the effect of $\delta_0$ and so we choose $\epsilon\sim {\cal O}(10^{-1})$ for a near-zone observer with $r_O \sim r_{\rm mid} \sim {\cal O}(10^3 r_S)$. This implies that for such near-zone observers, the effect of $\delta_0$ is about a hundred times larger than that of $\delta_{n\ge 1}$.

It is also worth mentioning that a more direct way of constructing the black hole shadows for a near-zone observer would be to adopt the numerical ray-tracing method in order to direct solve the geodesic equations. However, the near-zone metric \eq{sr_met_DY} holds only at ${\cal O}(\epsilon)$, and it is not easy to maintain this order of approximation when numerically solving the geodesic equations. It is also hard to estimate the order of accuracy. These difficulties can be controlled by the analytic approach, as discussed above.

We now adopt \eq{alpha-beta} to construct the shape of the shadow for the near-zone metric \eq{sr_met_DY} of a linearly superrotated black hole and a near-zone observer with $r_O \sim r_{\rm mid} \sim {\cal O}(10^3 r_S)$. Again, this is to emphasize the effect of the soft hairs with $\epsilon \sim {\cal O}(10^{-1})$. We first plug the expressions for $p^{\mu}$ given by \eq{pH_f} into \eq{alpha-beta}, and after some algebraic manipulations while keeping only the ${\cal O}(\epsilon)$ terms, the two shadow equations can be put into the following symbolic forms, respectively
\bea  
\alpha &=& ( -1+ \epsilon A_1) \lambda \csc\theta_O + \epsilon A_0 {\cal K}\;, \label{shape_1} \\
\beta &=& \epsilon B_2 \frac{\lambda^2 \csc^2\theta_O}{{\cal K}} + \epsilon B_1 \lambda \csc\theta_O + \; (1+\epsilon B_0){\cal K}\;, \qquad  \label{shape_2}
\eea
where $\cal K$ is given by \eq{R-K}. The $A_i$'s and $B_i$'s are independent of both $\lambda$ and $\kappa$ and are given by 
\bea 
A_0 &=& \sin\theta_O \partial_{\theta_O} Y^{\phi}\;, \\
A_1 &=& -\partial_{\phi_O}Y^{\phi}- 2 \partial_{\theta_O} Y^{\theta}\;, \\
B_0 &=& - \partial_{\theta_O} Y^{\theta}\;, \\
B_1 &=& -\csc\theta_O \partial_{\phi_O} Y^{\theta}\;, \\
B_2 &=& \cot\theta_O Y^{\theta}.
\eea
Here $Y^A=Y^A(\theta_O,\phi_O)$. For $\epsilon=0$, \eq{shape_1} and \eq{shape_2} yield a circular shadow described by $\alpha^2+ \beta^2=\kappa$, consistent with \eq{circle}.

In the case of nonzero $\epsilon$ and $Y^A$, we derive the shadow equation as follows. First, we solve \eq{shape_1} for $\lambda \csc\theta_O$, resulting in
\be 
\lambda \csc\theta_O = - \alpha +\epsilon \; \Big(-  A_1  \alpha+ A_0 \sqrt{\kappa - \alpha^2} \Big)\;, \label{sol_lambda}
\ee 
which leads to
\be 
{\cal K}=\sqrt{\kappa - \alpha^2} + \epsilon  \alpha \Big( A_0 -\frac{A_1  \alpha}{\sqrt{\kappa -  \alpha^2}} \Big)\;. \label{sol_cK}
\ee
From here, plugging \eq{sol_lambda} and \eq{sol_cK} into \eq{shape_2} and solving for $\kappa$, we obtain the shape equation of the shadow as follows
\begin{align}
 \Big(1- 2\epsilon (B_2-A_1) \Big) \alpha^2 &+ (1-2 \epsilon B_0)  \beta^2 - 2 \epsilon (A_0-B_1)  \alpha \beta = \kappa \;. \label{conicS}
\end{align}
In the above derivation, we have kept only ${\cal O}(\epsilon)$  terms as we only consider superrotation hair at the linear order \footnote{It is tempting to include the higher-order-in-$\epsilon$ terms to obtain more complicated and probably interesting shapes for the shadows, as \eq{shape_1} and \eq{shape_2} seem nonlinear in $\epsilon$. However, the terms of ${\cal O}(\epsilon^{n>1})$ are incomplete as we only include the superrotation hairs at ${\cal O}(\epsilon)$ in the metric \eq{sr_met_DY}.}.

\begin{figure}[H]
    \centering
 \includegraphics[scale=.6]{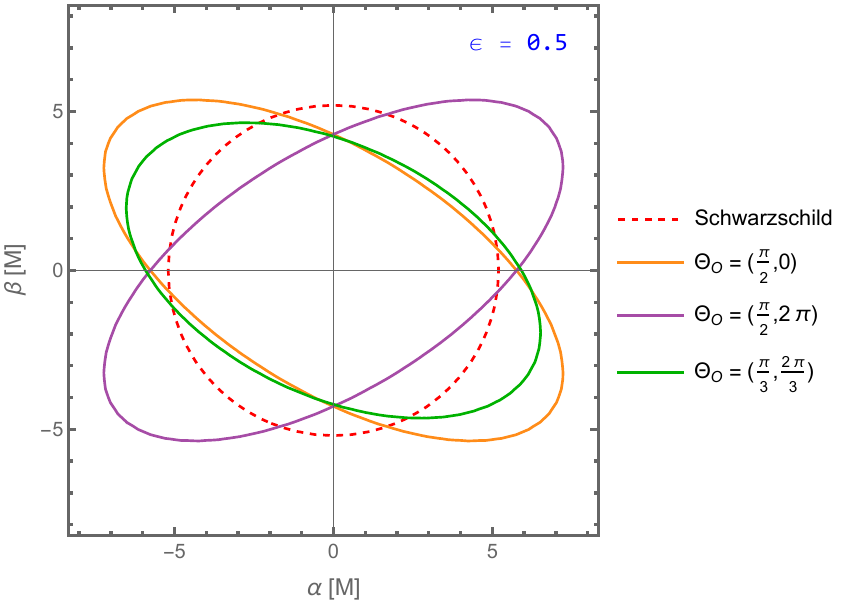}\vspace{.3in}

 \includegraphics[scale=.6]{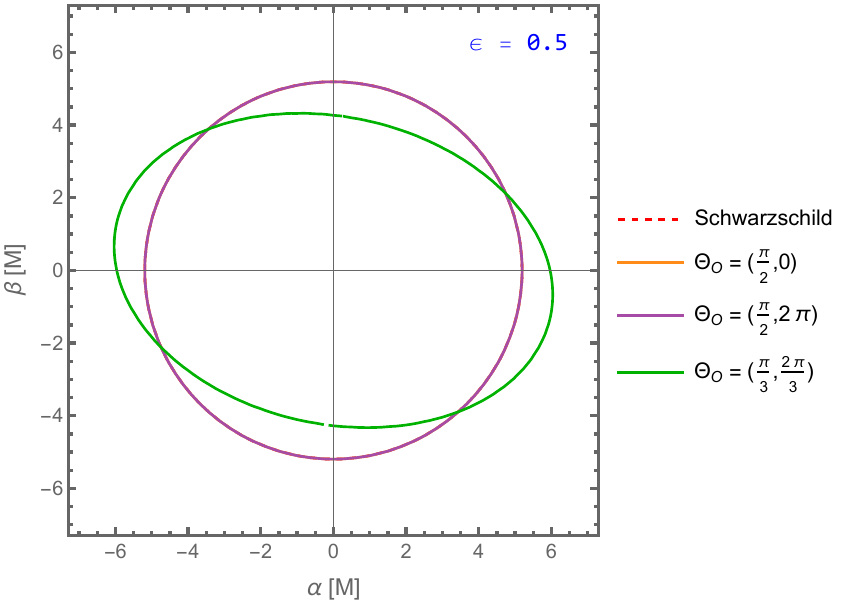}\vspace{.3in}

 \includegraphics[scale=.6]{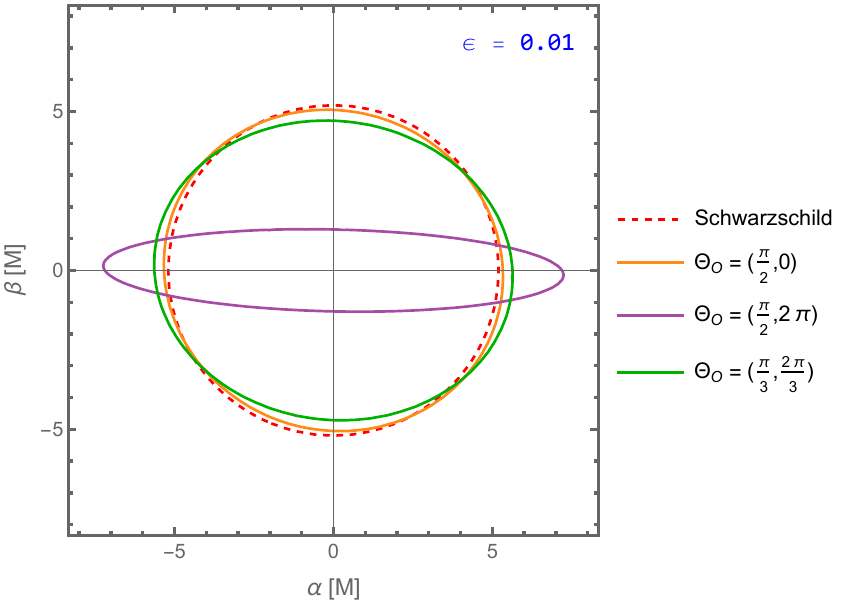}
 \caption{Examples of the shadows associated with the deformed light rings in Fig. \ref{fig1}, for three choices of observer each: $\Theta_O\equiv (\theta_O,\phi_O)=(\tfrac{\pi}{2},0)$ (orange), $(\tfrac{\pi}{2},2\pi)$ (purple), and $(\tfrac{\pi}{3},\tfrac{2\pi}{3})$ (green). The dashed red circle corresponds to the shadow of a bald Schwarzschild black hole. In the sinusoidal-type case (top), one sees that the observer's polar coordinate controls the overall size of the ellipse, while its azimuthal coordinate controls the orientation. For real spherical harmonics (center) the shadow tends to coincide with the bald case for equatorial observers, while deviating from this for inclined observers. The addition of a polynomial in $\theta$ and $\phi$ (bottom) causes some additional variation, which is more prominent near the azimuthal peak.} 
    \label{fig2}
\end{figure}

We see that the linear superrotation hair deforms the circular shadow of a Schwarzschild black hole into a conic section described by \eq{conicS}. In general, this conic curve is an ellipse and can be put into the canonical form,
\be
\frac{X^2}{\kappa a_+} +  \frac{Y^2}{\kappa a_-} =1
\ee
with 
\be
a_{\pm}= 1 \pm \epsilon \sqrt{(A_0-B_1)^2+(A_1-B_2+B_0)^2}
\ee
and 
\be 
(X,Y)^T = R(\Psi) \; (\alpha, \beta)^T,
\ee
where $R(\Psi)$ is a $2 \times 2$ rotation matrix of the $v_O$-independent angle $\Psi$ satisfying $\tan 2 \Psi = 2 \epsilon (B_1-A_0)$. The shape of the shadow depends on the observer's angular position $\Theta_O\equiv (\theta_O,\phi_O)$. Note that $A_1-B_2+B_0$ can be written as $-2 \partial_{\theta_O} Y^{\theta}$, and similarly $A_0-B_1=\csc\theta_O \big(\partial_{\phi_O}Y^{\theta}+\sin^2\theta_O \partial_{\theta_O} Y^{\phi}\big)$, by exploiting $D\cdot Y=0$. In Fig. \ref{fig2}, we show the shadows cast by the hair yielding the deformed light ring of Fig. \ref{fig1} for a few choices of the observer's position $\Theta_O$.

Some remarks on the conical shape of the shadow are in order. First, note that the discriminant of \eq{conicS} turns out to be $\Delta= 1$; thus, for black holes with near-zone linear superrotation hairs, the shadow takes the form of an ellipse. Second, the shape, size, and orientation of the shadow are time-independent, i.e., the lengths of major and minor axes, $\sqrt{\kappa |a_{\pm}|}$, and $\Psi$ are $v_O$-independent. Alternatively, the shape of \eq{conicS} can be characterized by its $v_O$-independent eccentricity $e$,  
\be 
e = \sqrt{1-\frac{a_-}{a_+} } = \sqrt{2\epsilon}\; \Big[(A_0-B_1)^2 + (A_1-B_2+B_0)^2 \Big]^{1/4} + {\cal O}(\epsilon^{3/2})\;. 
\ee
The center of the ellipse remains at the origin.

If a black hole is not far from an imaging telescope, one can detect the wide variety of elliptic deformations by linear superrotation hairs for such near-zone observers. However, this does not seem realistic in our current situation. Despite this, these results contrast sharply with the shadow of linearly supertranslated holes considered in \cite{Lin:2020gva, Zhu:2022shb}, which showed that the supertranslation hairs only cause non-observable static shifts of the shadow's center without any shape-changing \footnote{The methods used in \cite{Lin:2020gva} and \cite{Zhu:2022shb} are quite different. In \cite{Lin:2020gva}, a full numerical ray-tracing method is adopted to obtain the shadow for fully nonlinear supertranslated black holes. On the other hand, the method adopted in \cite{Zhu:2022shb} is the same as the one used in this paper applied to the case of linearly supertranslated black holes.}.  These deformations are also quite different from the possible shadow deformations caused by modified gravity because their patterns are fixed by only a few theoretical parameters and are thereby devoid of richness. In the future, a full metric with nonlinear superrotation hair is desired in order to explore the shadow deformation for far-zone observers via the numerical ray-tracing method. Once done, an EHT-like project can then be implemented to observe such soft hairs directly.

\section{Conclusion}\label{sec4}

In this paper, we first clarify issues regarding the singular pathologies of black holes dressed by soft hairs. We then focus on near-zone black hole metrics with linear superrotation hairs. These metrics are free of any near-zone singularities, which plague the far-zone metric with soft hairs. With such near-zone linearly superrotated black holes, we can construct the corresponding shadows observed by near-zone observers. Unlike the null effect of the supertranslation on the shadow as discussed in \cite{Lin:2022ksb, Zhu:2022shb}, we find that linear superrotation hairs can deform the circular shadow of a bald Schwarzschild black hole into that of an ellipse. Although this result is only for near-zone observers, this is the first evidence indicating that nontrivial infrared effects, such as soft hairs, can affect the black hole's shadow non-trivially.  To go beyond such a constraint, we need to patch the near-zone and far-zone metric with soft hairs, which is out of reach if no other ingredient exists to determine the matching condition. On the other hand, if the black hole solution with full nonlinear superrotation hairs is known, as in the supertranslated case, one can adopt the numerical ray-tracing method to obtain the shadows seen by far-zone observers.

In this work, we just focus on area-preserving near-horizon superrotations. By a similar procedure, one can obtain the corresponding effects for more general superrotations, i.e.,  those given in \eq{Rdecomp}. Unlike the gravitational memory effect associated with far-zone soft hairs, which causes a permanent shift of detectors' position, our finding on the effect of superrotations on the shadow yields a far richer observational signature.  
Furthermore, clarifying the singular pathology issues for soft hairs can help us to more firmly pin down the soft hairs' physical nature, as well as lift some conceptual misunderstandings, such as violating Birkhoff's uniqueness theorem, and removing any doubt of detecting the soft hairs via physical means. This also raises the issue of connecting the far-zone and near-zone metrics with soft hairs so that one can extend the scope of this work to middle-zone or even far-zone observers. This connection may also be related to the global topological nature of soft hairs and the possibility of realizing them as a topological defect. This work evokes all of these issues and deserves further study in the future.

\bigskip

\begin{acknowledgments}
We thank Ling-Wei Luo for his earlier participation in this project. FLL and JP are supported by Taiwan's National Science and Technology Council (NSTC) through Grant No.~109-2112-M-003-007-MY3 and 112-2112-M-003-006-MY3. AP is supported by NSTC grant No. 110-2811-M-003-507-MY2.
\end{acknowledgments}

\bibliography{references}
 
\end{document}